# Experimental phase control of a 100 laser beam array with quasi-reinforcement learning of a neural network in an error reduction loop.

Maksym SHPAKOVYCH[1], Geoffrey MAULION[1], Vincent KERMENE*[1], Alexandre BOJU[2], Paul ARMAND[1], Agnès DESFARGES-BERTHELEMOT[1], Alain BARTHELEMY[1].

(*) corresponding author: kermene@xlim.fr

Affiliations:

*(1) XLIM Research Institute, Université de Limoges-CNRS UMR n°6252, Faculté des Sciences et Techniques, 123 ave. A. Thomas, F-87060 Limoges, France*

*(2) CILAS Ariane Group, 8 avenue Buffon, CS16319, F-45063 Orleans cedex2, France*

**Summary:** An innovative scheme is proposed for the dynamic control of phase in two-dimensional laser beam array. It is based on a simple neural network that predicts the complex field array from the intensity of the induced scattered pattern through a phase intensity transformer made of a diffuser. Iterated phase corrections are applied on the laser field array by phase modulators via a feedback loop to set the array to prescribed phase values. A crucial feature is the use of a kind of reinforcement learning approach for the neural network training which takes account of the iterated corrections. Experiments on a proof of concept system demonstrated the high performance and scalability of the scheme with an array of up to 100 laser beams and a phase setting at $\lambda/30$.

## 1-Introduction

Recently, the number of applications requiring laser beam of high average power has strongly increased, addressing large projects such as space cleaning [1], spacecraft propulsion [2], particle acceleration [3], as well as industrial processes [4] or defense systems [5]. Laser beam combining is one of the most studied approach to reach very high power level, in particular the coherent beam combining (CBC) techniques [6]. They aim to phase lock the emission of a tiled laser beam array delivered by a network of amplifiers to generate a synthetic beam of high brightness. As the phase relationships between the beams in the array evolves over time in an actual laser system, especially in fiber laser system, these techniques have to correct the phase deviations from a synthetic plane wave in real time via a servo loop. CBC techniques have been widely developed in recent years, exploring different approaches to adjust the individual phases in the synthetic discrete wavefront. They can be classified in two broad categories. In the first one, the phase relationships of the beams in the array are measured and then corrected [7]. In the second one, the discrepancy between the actual wavefront and the desired wavefront is compensated in an iterative process [8]. In the latter case, an optimization algorithm drives the feedback loop, analyzing more global data on the array phase state from interference between all of the beams [9,10]. These techniques are often simpler to implement, with less electronic devices, at the expense of a more complex numerical processing and for some of them, at the cost of a lower speed for a large number of beams. This last issue is connected with the number of iterations required in the feedback loop to reach the expected phase chart which increases quickly with the number of phases to control. More recently, neural networks (NN) and machine learning were investigated in view of finding a potentially simpler and more efficient way for achieving coherent beam combining. One of the scheme covered in the published literature [11] relies on a direct phase recovery by a convolutional neural network (VGG) followed by a phase correction in one step, such as in the pioneering work on NN for adaptive optics [12]. The NN serves to map the intensity of an interference pattern of the beam array (far field formed at a lens focus or an image out of the focus, power behind a beam splitter, etc.) directly into the distribution of phase in the array. Once the initial phase map is recovered, it is straightforward to apply phase modulation to set the phases to the desired values. The simulations reported in [11] show that the accuracy of the CNN based phase control drops when the array increases from 7 to 19 beams. This is a limitation which was also highlighted in the field of wavefront sensing so that NNs were often used only as a preliminary step for initialization of an optimization routine [13]. Another possible scheme is reinforcement learning. It was investigated experimentally and applied to the basic situation of the coherent summation of two beams from a fibered Mach-Zehnder interferometer [14]. In that case, the network (DQN) learns how to optimize a reward parameter (the combined power) which should be maximized in the desired phase state. Then, once trained, it commands directly the differential phase compensation. It was shown that the network yields phase noise control with efficiency and speed but scalability remains questionable, in particular in view of the huge duration of the training period even for two beams only.

The scheme we propose constitutes a third approach where accuracy and speed are compatible with scalability. It is first validated by numerical simulations and then experimentally in a proof of concept experiment with up to 100 laser beams.

## 2- Neural network in an error reduction physical loop with a specific reinforcement learning

The basic principle of the proposed scheme is as follows. Since the accuracy of NN based direct phase recovery significantly and quickly decreases when the number of waves to control grows, we suggest to include the network into an iterative process of error reduction in order to get phase-locking on any desired phase chart, in a progressive way after a few steps of phase modulation. The first idea could be to include a neural network designed and trained for direct (one step) phase recovery in a looped system with a phase modulator. However, the convergence was not observed in our computations and increasing the number of iterations in the loop does not help to improve this feature either. That is the reason why we specifically trained the network to operate in an error reduction loop, with a given small number of iterations T, using a simplified form of reinforcement learning technique. The phase control architecture is schematically depicted on Fig.1. The individual amplitudes of the laser fields are known and almost uniform, $|z_k| \approx 1$, but their phase is unknown. For that reason, the current state of the laser fields array is analyzed by diffraction after transmission by a diffuser. The scattered intensity pattern depends on the phase distribution in the array. It is measured by only a few photodetectors sparsely distributed in the scattered image. Their output data feed the NN that gives the corrections values which are then applied on the phase modulators. The process is iterated $T$ times.

All these elements, NN in a physical loop and reinforcement learning adapted to an iterative process, make our proposition innovative and unique. It is also possible to formulate our approach in terms of learning recurrent neural network, however a slight modification of reinforcement learning framework fits more natural for our aim.

In practice, in a situation where the beam array phase distribution to control is no longer static but evolves continuously over time, because of environmental perturbations and noise, the feedback loop for phase correction is kept permanently in operation. So in a dynamic situation the number of iterations in the loop is no longer fixed except during the training phase.

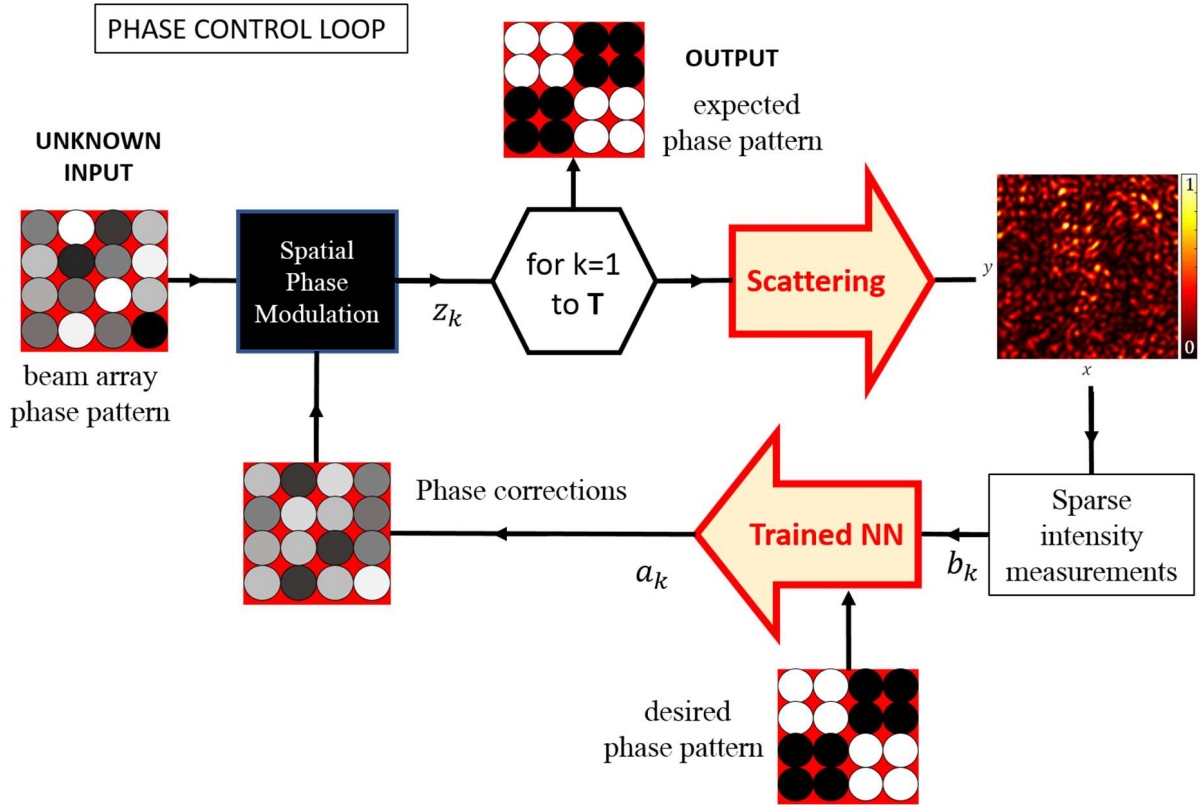

*Figure 1 : Schematic depicting the architecture of the laser array phase control system in a particular example of desired phase pattern with four zones of 0 and $\pi$ values in a 4x4 beam square array. (The phase values are encoded here between 0 and $\pi$ in a grey scale).*

### *Specific quasi-reinforcement learning*

Generally speaking, reinforcement learning is learning by interacting with an environment which rewards action made by an agent. Basically, from an observable environment state, the agent chooses to take an action in the environment for which it is rewarded accordingly. The way in which the agent chooses action is called a policy. Obviously, the agent aims at increasing the reward it receives and so must learn an optimal policy for interacting with the environment [15, 16].

In our particular case, the environment at the $k^{th}$ step, consists in the $n$ laser fields in the array $z_k \in \mathbb{C}^n$, the user-defined desired signals $z_d \in \mathbb{C}^n$ and the vector of scattered intensity measurements $b_k \in \mathbb{R}_+^m$, where $m$ is the number of detectors ($m > 2n$). The agent's policy is a neural network, which has to be designed and trained. The observable environment state is the vector $b_k$. The action $a_k \in \mathbb{C}^n$, as it is concerned, is a signal correction resulting from the agent's policy whom relevance is assessed from the reward $r_k$. For that purpose, we chose as a reward the following resemblance parameter

$$R(z_k, a_k) = \frac{|\langle z_k, a_k \rangle|^2}{\langle |z_k|, |a_k| \rangle^2},$$

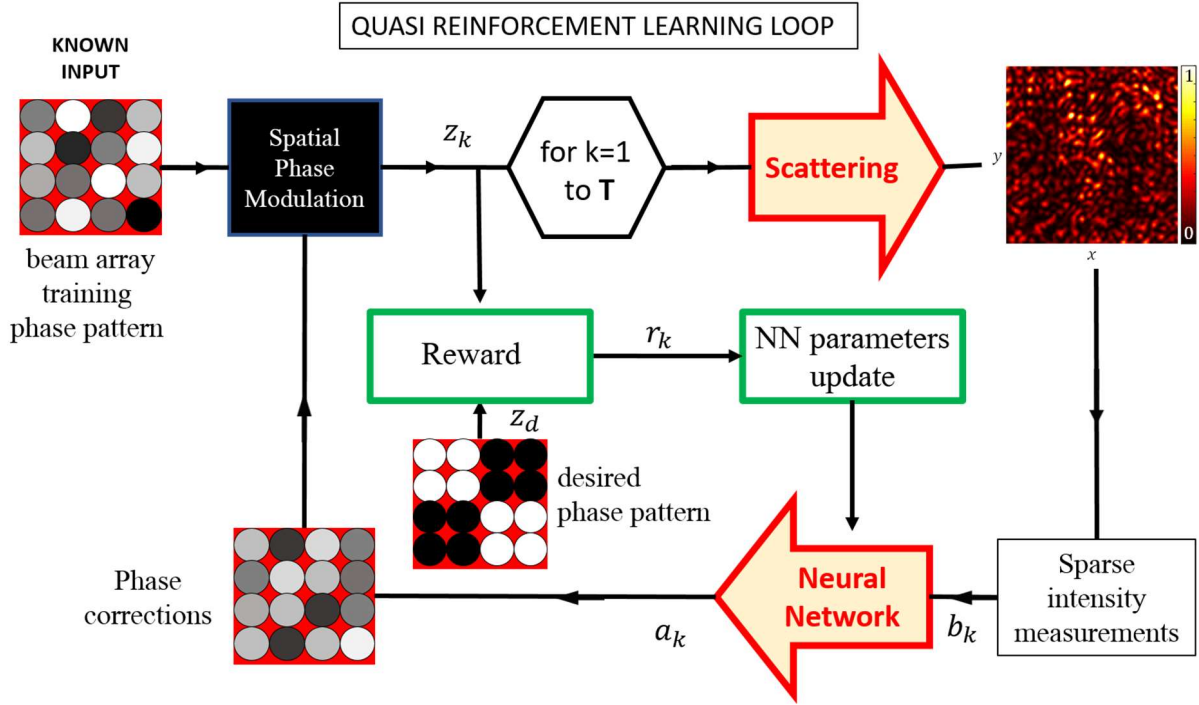

*Figure 2: Schematic of the specific reinforcement learning loop used for each phase pattern of the training data set, with a particular example of desired phase pattern.*

which is usually named phasing quality in the context of laser coherent beam combining. The reward is maximum and equals one if and only if $\arg(a_k) = \arg(z_k)$ up to a constant phase. Finally, the agent's interaction with the environment, since the action is just a phase correction, can be expressed as

$$z_{k+1} = z_k \cdot e^{i(\arg(z_d) - \arg(a_k))}.$$

As mentioned before, reinforcement learning approach proposes to find the agent's policy from the parametric family of neural network functions. The main difficulty in the general case for reinforcement learning is that we do not know the correct action $a_k$ at state $b_k$. This fact leads to the different approaches of learning agent's policy, which are all based on the repetitive collection of a huge number $N$ of action-state-reward triplets $\{(a_k, b_k, r_k)\}_{k=1}^{N}$ that are used to optimize NN's parameters. However, in the present phase correction problem, correct actions can be known during training and as a consequence, reward $r_k$ can be calculated in the same way as a loss function for supervised learning. In effect, during the learning process, known array phase patterns feed the process, so that $z_k$ is known and so that reward can be computed $r_k = R(z_k, a_k)$. We observed that if the agent is trained to maximize the reward **at each iteration**, for a fixed total number of iterations $T$, then actions $a_k$ are such that $\lim_{k \to \infty} \arg(z_k) = \arg(z_d)$ up to a constant. To emphasize the fact that it is not the classical case, we call it quasi-reinforcement learning (QRL). A simplified picture of the learning scheme is given in Fig.2.

In practice, in our simulations, to achieve a wavefront setting with a rms accuracy $\leq \lambda/30$ ( $r_{k_{\max}} \geq 0.96$) [17], $T$ must be greater than a critical value (typically 4 to 8) which depends on the number of beams $n$ and on the number of measurements $m$. More details are provided in the next section.

Note, we can either predict a phases vector or, directly, real and imaginary parts to build an action $a_k$. We found that the prediction of the real and imaginary parts of a signal instead of exact phase values allows the network to learn better, keeping in mind that modulus are known.

## 3-Simulations

Learning of the NN was made with up to 1000 epochs of 1024 random samples (generated for each learning step) of laser field array ($z_k$) with their associated scattered intensity ($b_k$), the half of which served for the training and the remaining served for tests. The samples can be experimentally obtained by sending laser fields array with known phase patterns on the diffuser and by recording the associated measured intensity. In order to speed up the training data generation, the environment can be represented by a complex valued transmission matrix TM or by another neural network NN-G, like done for image transmission through multimode fiber [18]. Both options work well and require far less experimental data. Once the TM or the NN-G is known, it is then fast to generate numerically any batch of training data for learning the NN to be used in the phase correction system. Optimization of the NN parameters was achieved with the Adam optimizer [19] (default parameters from the original paper) using for loss function $L$, such that $L(x,y) = 1 - R(x,y)$, $R(x,y)$ being the reward function. Computations were carried on a computer under Ubuntu 18.04 OS with GPU - NVIDIA GTX 1050, CPU - Intel Core i7-8750H and RAM - 16GB DDR4. We investigated by simulations the impact on the mean value of the phase control level $\overline{r_T}$, (i) of the NN architecture, (ii) of the number $T$ of iterations in the loop and (iii) of the number $n$ of laser beams in the array. The parameter $m/n$ was kept fixed and equal to 6 in the reported results.

The study indicated first that a simple NN with one hidden layer (a perceptron) is the most efficient structure. More hidden layers, activation layers or even convolutional layers did not evidence better operation. For the selected NN, Fig. 3 shows various evolutions of the mean reward $\overline{r_T}$ which denotes the reward at the last correction, averaged on a batch of 512 test samples. In Fig. 3-(a), for the case of 32 beams, the plots report the mean reward according to the training epoch for $T \in \{2, 4, 6, 8\}$. As mentioned above, it shows in this example that $T$ must be equal to or greater than 4 if a high accuracy ($\leq \lambda/30$) is required. In Fig. 3-(b), the traces present the mean reward according to the training epoch for various number of beams $n \in \{16, 32, 64, 128\}$ and for a fixed number of phase corrections $T = 6$. It is not surprising to see that the training takes more epochs when the size of the array to be controlled increases, varying from 50 to ~ 1000 when n grows from 16 to 128. That is connected with the number of parameters to optimize in the NN which scales as $4n(m + 2n + 3/2)$.

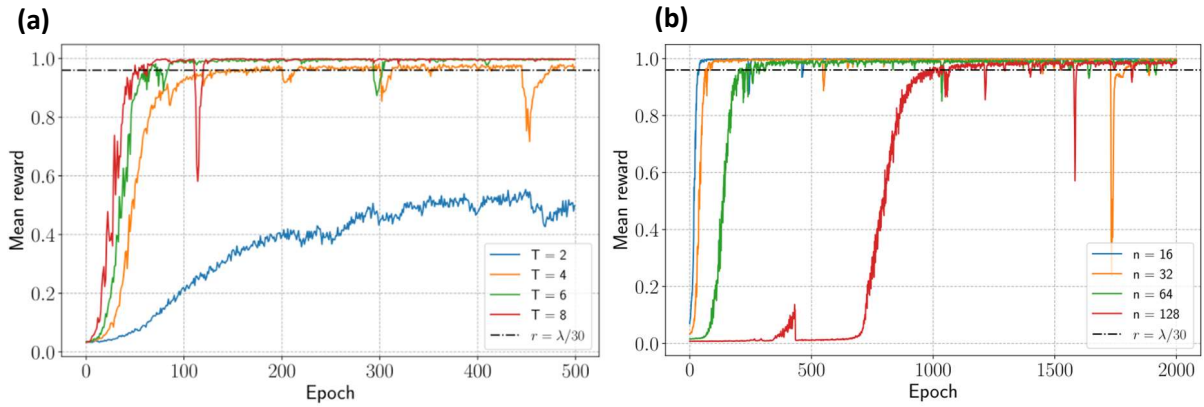

*Figure 3: (a)- Reward evolution during learning process for a fixed 32 beams example and different number of corrections and (b)- for a fixed T= 6 corrections and different number of beams. The black horizontal dashed line corresponds to a phase control accuracy of $\lambda/30$ rms.*

Fig. 4 depicts the time in seconds that is required for learning the network for $n \in \{16, 32, 64, 128\}$ and $T = 6$, where the stopping criterion was an achievement of $0.96$ value for the reward $r_T$. The evolution is well fitted by a parabola. Increasing either the number of detectors or the number of corrections will not affect the time dramatically because of parallel computation on GPU. The capacity of used GPU is high enough to compensate for the increase of parameters and keep approximately the same computation time.

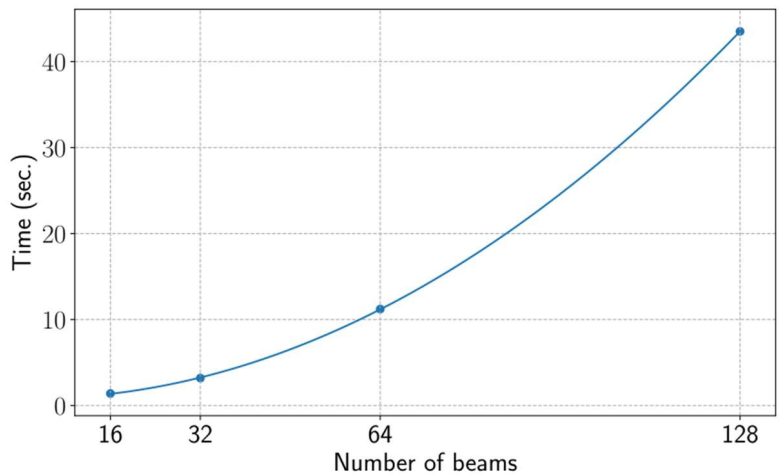

*Figure 4: Required time for training the network up to a reward of 0.96 for different number of beams. Dots stand for the measured training time and the curve is a fit by a parabola.*

One can see in Fig. 5-a the reward changes due to a variation in the number of iterations in the loop $T$, for the case of 32 beams, and in Fig. 5-b, the changes when the size of the laser array grows from 16 up to 128 beams for $T = 6$. These results show that about 3 corrections could be sufficient to get a phase control down to $\lambda/30$ accuracy or better for $n = 16$ and 6 iterations for $n = 128$, even if the precision still improved for a larger number of round in the loop. It demonstrates as well that the scheme is scalable without losing much in performance.

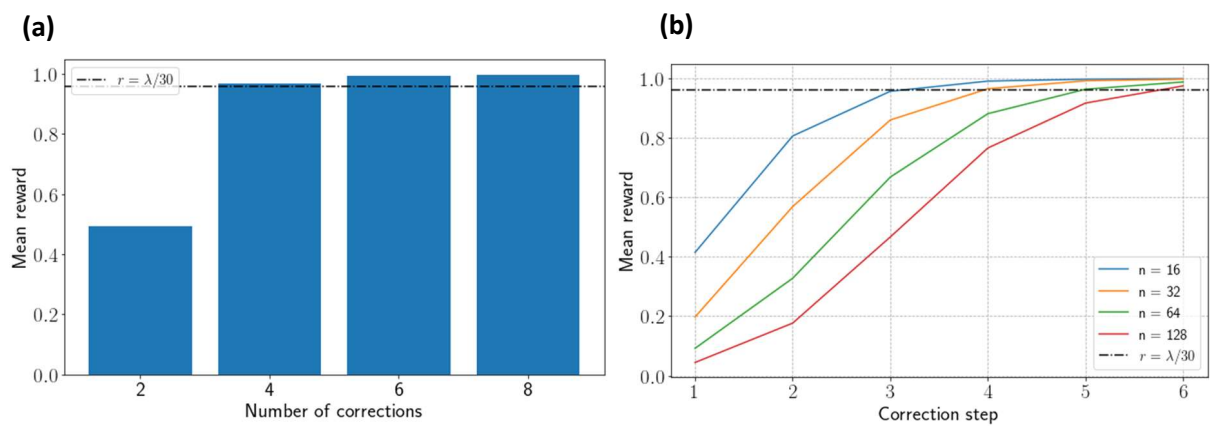

*Figure 5: (a)- Evolution of the final reward for n=32 when the number of corrections changes (m/n=6). (b)- Reward evolution in correction loop for different number of beams. The black dashed line corresponds to a phase control accuracy of λ/30 rms.*

## 4-Experiments

The experimental set-up used to make a proof of concept experiment, is schematically represented on Fig. 6. A first part of the set-up served for the beam array generation. The beam from a 1064 nm fiber coupled laser diode was collimated, linearly polarized, and then expanded by a telescope with 5.6 magnification. The enlarged beam (ECB) was launched on a metallic mask (M) with many circular apertures drilled according to a square array. The transmitted waves formed the laser field array which size can be varied from 16 beams up to 100. Next, they were imaged by a 1:1 telescope (L1, L2) on a first reflective spatial light modulator (SLM1) which allowed an individual control of the beamlet phase in the array. To be more precise, it is the individual piston phase which is under concern here, assuming that each elementary beam can be considered as a plane wave. Thus, the reflected beam array mimics the output of an array of parallel laser amplifiers with single mode output. It is further imaged by a second telescope (L3, L4) on a second phase modulator (SLM2) for phase control of the final beam array delivered to the application. The output is split in two parts by a non-polarizing beam splitter (NBS). One fraction was sent on a ground glass diffuser to get interferences between the optical fields in the whole array. The second output of the NBS was focused by a positive lens (L6) in order to observe the far field intensity pattern of the beam array with a camera (Cam2) located in its back focal plane. The interferences were observed as a speckle pattern after transmission of the beams through the ground glass plate (GGP) and diffraction in free space on a few centimeters. The speckle intensity image was detected in a sparse way, in a few transverse positions only, by an array of photodetectors (here some pixels in a camera sensor Cam1). The measured intensity is digitized and then transmitted to the computer for processing by the neural network. The NN gives the phase corrections required to set the laser fields closer to the desired phase values. The phase corrections are thus transmitted to SLM2 for modulation and correction of the input optical beams, which starts a new round in the loop until a steady state is reached. Our experimental study covers different beam array size, from 4x4 up to 10x10, as well as different number of intensity sample in the scattered pattern, m/n ratio ranging between 2.2 and 7.5. However, we chose to only report in the following the results obtained with 100 beam array, the greatest number we could achieve in practice with our set-up. To prepare the learning stage, 1000 probe beam arrays with various random phase pattern were consecutively launched on the diffuser and the corresponding scattered intensity were detected and recorded. This set of experimental data served to get the complex values transmission matrix (TM) of the diffuser by use of the optimization method presented in [20].

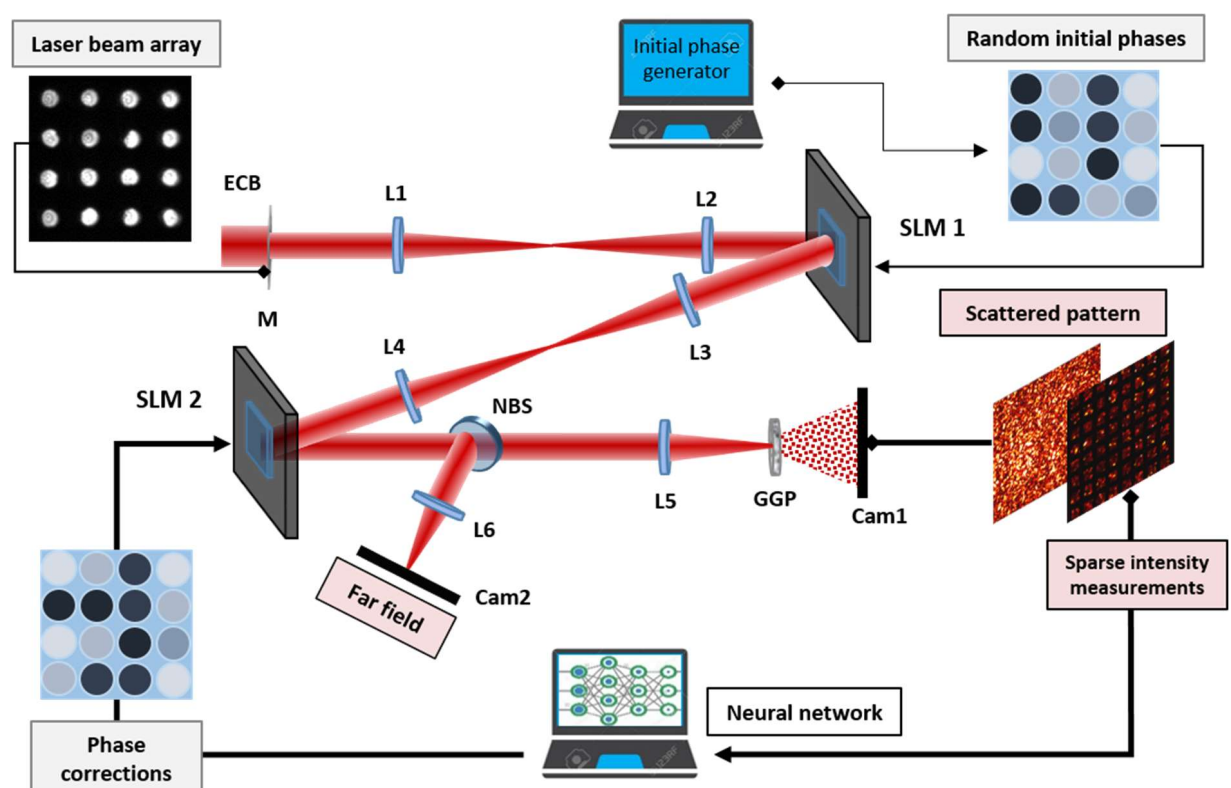

*Figure 6 : Experimental set-up for the proof of concept experiments. A broadened laser beam (ECB) from a laser diode was filtered by a mask with an array of hard apertures (M) and the phase of the beam in the array was further scrambled by reflection on a spatial light modulator (SLM1). This part served for the generation of beam array with random phase states mimicking an input with unknown phase map. The phase control loop starts with a second phase modulator (SLM2) for phase correction. The corrected beam array was focused on a ground glass diffuser (GGP) giving a speckle pattern carrying the phase information. The scattered imaged is measured in some sparsely distributed position by selected region of interest in a camera sensor (Cam1). The data were transmitted to the QRL-trained neural network which delivered the phase modulation to apply on SLM2 for appropriate convergence to the desired phase map. Observation of the output beam array far field in the focus of L6 by means of a second camera (Cam2) permitted a complementary assessment of the phase control operation.*

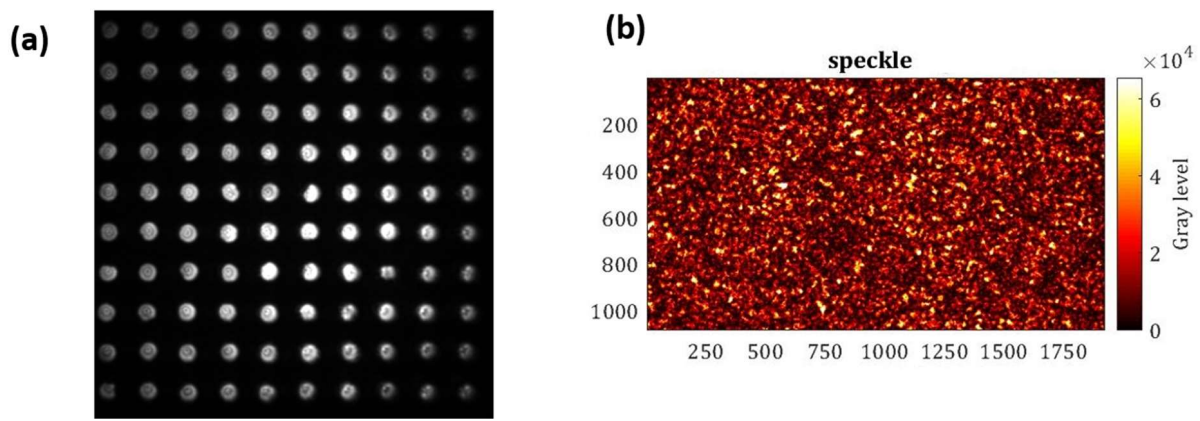

*Figure 7 : (a) Image of the array of 100 coherent laser beams in a square lattice we used in experiments. (b) Typical scattered pattern observed behind the diffuser where an array of a few photodetectors measure the optical intensity on some sparsely distributed transverse positions only.*

The TM was further utilized to compute batches of 512 couples of input/output training data which were varied for each training epoch. Then the NN was trained by quasi-reinforcement learning like done in the simulation section with a parameter $T$ set to 6. A typical evolution of the reward during the quasi-reinforcement learning process (training epoch) is shown on Fig.8 in the case of a 10x10 beam array. A plateau close to the reward value of 0.99 is reached here after about 500 epochs. The blue trace corresponds to the training data set and the red trace to the test data set. They coincide almost perfectly. The training lasted about 8 minutes on a simple laptop.

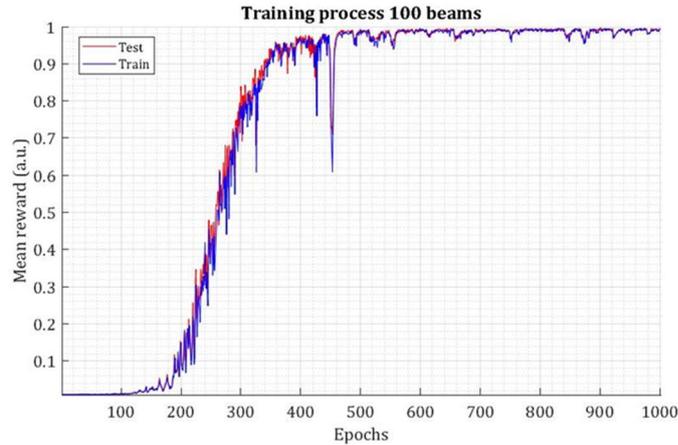

*Figure 8: Evolution of the reward parameter averaged over a batch of 500 tests versus training epochs, with training experimental data for a 100 beam array*

Then the QRL trained neural network was implemented on the computer driving the phase correction performed by the SLM in order to assess the operation of the laser array phase control system. For each test, the beam array was kept with the same almost uniform intensity (see Fig.7-a) and a random phase pattern was chosen as input of the system. The desired final phase distribution was selected as uniform as required in the case of standard beam phase locking for coherent combining. The phase control system was started and operated as a feedback loop for 20 rounds. At each round we computed the correlation between the expected phase distribution and the current state, in other words we computed the phasing quality. The evolution of the phasing quality versus the correction steps ($\equiv$round in the loop) were gathered for 100 different tests which are plotted on Fig 9 in the case of a 100 beams square array (10x10). A trace of the average evolution is shown in red. In all cases, the experimental phasing quality quickly raised from the low initial state to a high value at steady state. On average it takes about 6 rounds (6 corrections) to reach a phasing quality of 0.96 which corresponds here to a discrete wavefront with less than $\lambda/30$ deviation (rms) from a plane wave. These experimental values, which are in excellent agreement with the training and with the simulations, evidence the efficiency of the new learning based phase control system. To our knowledge, it is the first experimental results with such a high number of beams for a laser phase control device operated with a neural network.

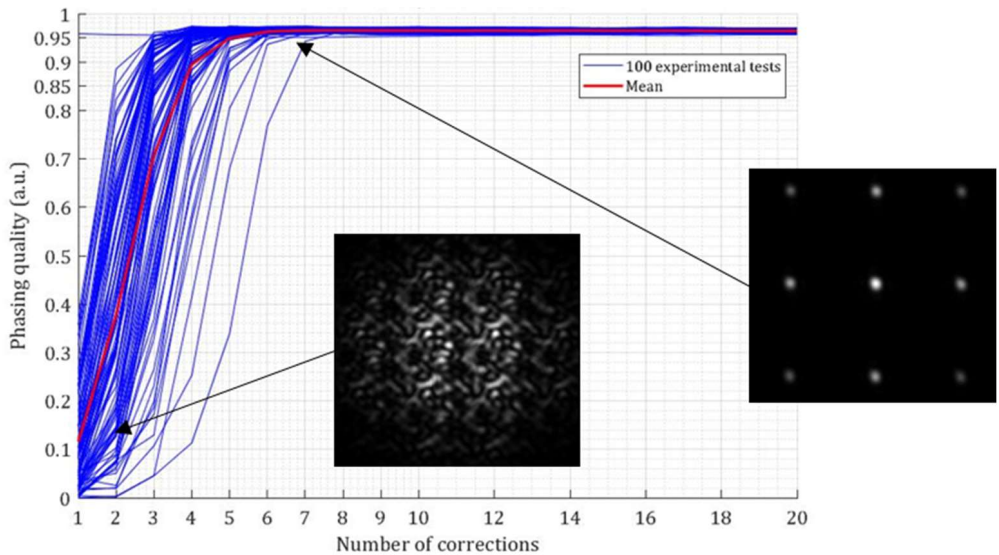

*Figure 9 : Phasing quality owing to the number of correction steps in a 100 beams co-phasing experiments. Each of the 100 plots in blue corresponds to a test initiated with a different random phase chart. The average trace is shown as a red line. The phasing quality reaches its maximum at steady state in 6 corrections in average with a value of 0.96. Insets corresponds to far field intensity images of the beam array recorded by Cam2 at the initial state and after six corrections where phase locking was achieved.*

**5-Discussion**

For every particular situation regarding the laser beam array, the performance of the phase control system will depend on the array size n, and on the parameters $T$ and $m/n$. For a fixed value of $m/n$, the number of corrections steps required to get the same high threshold accuracy $T_{th}$ grows when the number of beams to control increases. Nevertheless the evolution is slow and is well fitted by a logarithmic law, $T_{th} = \log_2(n) + cst(m,n)$ ($cst \leq 0$), where the constant depends logarithmically on the ratio $m/n$ starting from $cst$ =0 for $m/n$ =4. So for a twofold increase in the array size only one additional correction is needed. For a fixed array size n, an increase in the number of measurements m leads to an accuracy improvement for a fixed number of corrections or alternatively it leads to a lower number of corrections to reach a given phase setting accuracy. On one hand, a fast system would require a small $T$ because the bandwidth scales as $(T.\tau)^{-1}$ where $\tau$ is the duration of one loop. A low cost system would call for a small $m/n$ in order to use the lowest number of photodetectors and their associated electronics. On the other hand, a high accuracy would call for a high value both for $T$ and $m/n$. In the case of 100 beams for example, two options give similar accuracy: $T$ =6 and $m/n$ =6 or $T$ =8 and $m/n$ =4. Therefore, a trade-off has to be found to get a fast accurate system at the lowest price.

In our proof of concept experiments the time required to operate one loop of the system was given by the actuation speed of our liquid crystal based phase modulator (SLM) and the non-optimized recording-reading steps of the intensity sensor. One round in the loop took about 400 ms. On an actual fiber laser array, the phase modulators are usually fast electro-optic modulators with tens of GHz bandwidth so that speed limitation would still come from the reading speed of the photodetectors used for the scattered intensity measurements. With a specific design, an array of individual photodetectors could have a 1 MHz bandwidth but arrays of detectors are usually slower. It is worth mentioning that the measurements of light intensity in the scattered image do not need to sample the speckle pattern according to its two transverse dimensions. As it was demonstrated experimentally with our set-up, the sparsely distributed measurements can be done along a straight line across the scattered image without losing the information required for operating the phase control. That would permit the use of cheap linear detector array which reading speed is fast (~100 kHz).

In simulation and in experiments the observed performances did not evolve if the desired pattern was changed from a uniform phase map, like for coherent combining, to a more structured map.

## 6-Conclusion

We have proposed a new scheme for the phase control of a coherent laser beam array such as the ones encountered in coherent beam combining. It is based on a feedback loop including sparse intensity detection of a scattered pattern of the beams followed by a specifically trained neural network for derivation of the phase corrections to apply to the phase modulators. Novelty stands in the fact that the system is designed and trained to operate in a loop with a fixed low number of iterations. The neural network is trained by quasi-reinforcement learning, a specific simplified form of reinforcement learning. By comparison with the state of the art of learning based approaches for CBC, the new scheme features a faster training, a high accuracy and most of all an unprecedented potential for scalability. Beyond validation of the proposed technique by numerical simulations, a proof of concept experiment demonstrated the efficient phase control of a 100 beam square array with performances in agreement with the simulations. Scalability preserves accuracy and the reduction in terms of bandwidth is weak since it scales as $(\log_2(n))^{-1}$. The obtained results establish a new record, both in simulations and in experiments, for phase control of beam array based on learning techniques. This approach sounds promising for directed energy applications and for fiber laser amplifier array.

**Acknowledgments**: The first author was supported by institutional grants from the National Research Agency under the Investments for the Future program with the reference ANR-10-LABX-0074-01 Sigma-LIM, and by the European Regional Development Fund and Nouvelle Aquitaine Region.